\documentclass[apl,numerical,reprint,showpacs]{revtex4-1}

\usepackage{graphicx}
\usepackage{bm}
\usepackage{color}
\usepackage{textcomp}

\def\um{\hbox{\textmu}m}
\def\uA{\hbox{\textmu}A}
\def\nm{\rm{nm}}
\def\Ic{I_{\rm{c}}}

\renewcommand{\deg}{^\circ}

\begin{document}

\title{Critical-Current Reduction in Thin Superconducting Wires Due to Current Crowding}

\author{H. L. Hortensius}
\affiliation{%
	Kavli Institute of Nanoscience, Faculty of Applied Sciences, Delft University of Technology, Delft, Netherlands}

\author{E.F.C. Driessen}
\affiliation{%
	Kavli Institute of Nanoscience, Faculty of Applied Sciences, Delft University of Technology, Delft, Netherlands}

\author{T. M. Klapwijk}
\affiliation{%
	Kavli Institute of Nanoscience, Faculty of Applied Sciences, Delft University of Technology, Delft, Netherlands}

\author{K. K. Berggren}
\altaffiliation[also:~]{Research Laboratory of Electronics, Massachusetts Institute of Technology, Cambridge, MA, U.S.A.}
\affiliation{Kavli Institute of Nanoscience, Faculty of Applied Sciences, Delft University of Technology, Delft, Netherlands}

\author{J. R. Clem}
\affiliation{%
   Ames Laboratory and Department of Physics and Astronomy, Iowa State University, Ames, Iowa, 50011 USA}

\date{\today}


\begin{abstract}
We demonstrate experimentally that the critical current in superconducting NbTiN wires is dependent on their geometrical shape, due to current-crowding effects. Geometric patterns such as $90\deg$ corners and sudden expansions of wire width are shown to result in the reduction of critical currents. The results are relevant for single-photon detectors as well as parametric amplifiers.
\end{abstract}

\maketitle

Superconducting wires made of strongly disordered superconducting materials such as NbN and NbTiN are used for single-photon detection\cite{Goltsman,Kerman,DorenbosSPD}, single-electron detection\cite{Rosticher} and parametric amplification\cite{Eom}. In all cases, for optimal performance, the devices are biased at as high currents as possible, without exceeding the critical current. In principle, for wires smaller than both the Pearl length $\Lambda = 2\lambda^2/d$ ($\lambda$ is the dirty London penetration depth and $d$ is the film thickness)\cite{Pearl} and the dirty-limit coherence length $\xi$, the critical current is determined by the critical pair-breaking current, which has been theoretically calculated over the full temperature range by Kupriyanov and Lukichev\cite{Kupriyanov}. These predictions have been tested experimentally in aluminium by Romijn et al.\cite{Romijn} and Anthore et al.\cite{Anthore}. In many practical cases, it is found that the critical current varies from device to device and is significantly lower than this intrinsic maximum value. This reduction is usually attributed to defects in the films and slight variations in width. In addition, for strongly disordered superconductors electronic inhomogeneity may develop, even for homogeneously disordered materials\cite{Sacepe}. However, Clem and Berggren\cite{Clem}, responding to the observed dependence of the critical currents of superconducting single-photon detectors on the fill factor of the pattern\cite{Yang}, explained that the critical current may depend on geometric factors in the wires, such as bends. In their model analysis, the superconducting wires were narrower than the Pearl length, but wider than the coherence length. Consequently, the current is not necessarily uniform and the critical current is reached when the current density locally exceeds the critical pair-breaking current. At this current a vortex enters the superconducting wire, causing the transition to a resistive state\cite{Bulaevskii}. In this letter, we present an explicit comparison of critical currents of superconducting NbTiN nanowires with different geometrical shapes and confirm that the observed critical current depends on the geometry.

\begin{figure}
\includegraphics{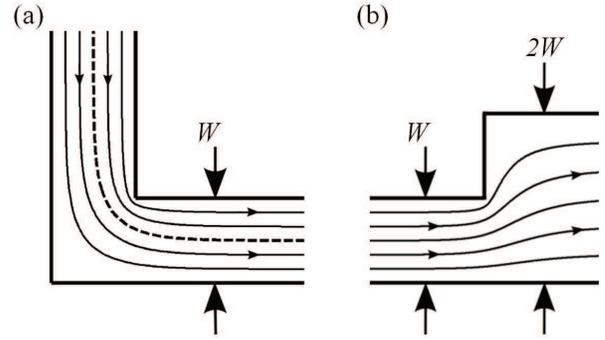}
\caption{Calculated current stream lines for the cases of (a) a $90\deg$ bend and (b) a suddenly widening ``stub'' region. The increased density of stream lines near the inner corners shows the current crowding. The dashed curve in (a) represents the median contour: half the integrated current flows on one side of the contour, and half flows on the other side. An inner corner shaped according to this contour will exhibit no current crowding, and is said to be optimally rounded.}
\label{rtangleplot}
\end{figure}

In the analysis of Clem and Berggren\cite{Clem}, the sheet-current distribution is calculated for a given geometry using conformal mapping. Then, the critical current is defined as the current at which the Gibbs-free-energy barrier for a vortex to enter the film is reduced to zero. Any geometry where the current is led around a sharp feature, even when spreading into a suddenly widening portion of the wire, is predicted to lead to an increase in local current density at the inner curve of the corner, and a correspondingly reduced critical current. This behavior is similar to that in a normal-metal current divider, in which the shorter, lower-resistance, path around the inner corner carries a higher proportion of the current than the outer, higher-resistance, path. Fig.~1 shows the current streamlines in a superconducting thin-film wire for two cases: (a) a sharp $90\deg$ turn and (b) a sudden widening or stub. Note the increased current density in the inner corner. The dashed curve in (a) follows a contour where the current density equals the current density of an infinite straight wire. Outside this contour, the current never exceeds the average current density in the straight wire segment. Therefore, a corner conforming to this geometry will always be able to carry the current supplied to it, without exhibiting a reduction in the critical current. The shape of this optimally rounded corner is given in Eqs. (112) and (113) of Ref.~\cite{Clem}. Even though in the structure in Fig.~1(b) material is added with respect to a straight wire of width $W$, the critical current is still predicted to be reduced.

\begin{figure}
\includegraphics{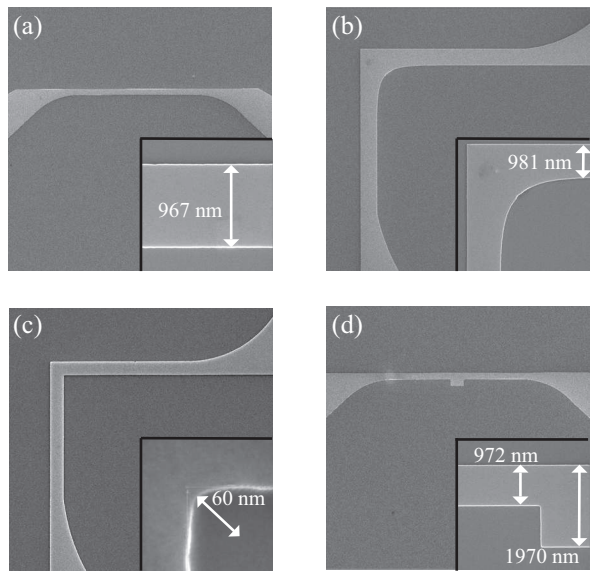}
\caption{SEM images of test structures measured in these experiments: (a) a straight line segment; (b) an optimally rounded $90\deg$ bend; (c) a $90\deg$ bend; and (d) a suddenly widening ``stub'' region. The minimum inner radius of curvature achievable in our fabrication process was approximately 60\,\nm. The light spot visible on (d) is due to charging after zooming in on the wire-contact transition region.}
\label{SEMs}
\end{figure}

Fig.~\ref{SEMs} shows examples of the four patterns we used to test the theoretical predictions of Ref.~\cite{Clem}: (a) straight wires (3 devices); (b) corners with optimally designed inner curves (2 devices); (c) corners with sharp inner curves (3 devices); and (d) a straight line that included a wider region (a stub) extending 1\,\um~to one side (1 device). All geometries consisted of a 1-\um-wide wire with a length of 20\,\um. The devices were contacted at their terminals by using a gradually widening region, to avoid potential current crowding at the contacts.

To quantify the expected effect, we define the reduction factor $R$ to be a number between zero and one, where the critical current of the geometry is given by $\Ic = R\cdot I_{\rm c,straight}$, with $I_{\rm c,straight}$ the critical current of an infinitely long wire of width $W$. $R=1$ for the straight wire and for the optimally rounded corner, indicating an unsuppressed critical current. For the sharp corner and the wire with a stub, $R\propto(\xi/W)^{1/3}$, with $\xi$ the coherence length and $W$ the width of the wire. The suppression is thus larger for shorter coherence length and larger wire width. In order to satisfy the condition $\xi \ll W \ll \Lambda$, we fabricated the wires of width $W$ = 1\,\um~in a~8\,nm thick NbTiN film with a resistivity of 170\,$\mathrm{\mu\Omega cm}$, implying a mean free path $l \approx 0.6$\,nm. The coherence length $\xi \approx \sqrt{\xi_0 l} \approx 7\,\mathrm{nm}$, with $\xi_0$ the BCS coherence length, and the Pearl length $\Lambda \approx 20$\,\um.

The NbTiN film was grown on a high-resistivity ($\rho > 1\,\mathrm{k\Omega cm}$), hydrogen-passivated silicon substrate using DC magnetron sputtering from a 70\% Nb, 30\%Ti target, at a pressure of 8\,mTorr, a DC power of 300\,W, a nitrogen gas flow of 4\,sccm and an argon gas flow of 100\,sccm. Onto this film, a 150\,\nm-thick layer of hydrogen silsesquioxane resist was spin-coated. This layer was patterned using electron-beam lithography (100\,keV, $\sim2$\,mC/cm$^2$). The resist was developed in a tetra-methyl ammonium hydroxide solution and a post-development exposure was performed to harden the resulting etch mask\cite{YangPostExp}. Finally, the pattern was transferred into the NbTiN film by reactive-ion etching in a 50\,W SF$_6$/O$_2$ plasma, using interferometric end point monitoring. Both a large-area test structure and the wires showed a critical temperature of 9.5\,K.

Devices were inspected using electron microscopy after electrical testing. All structures designed to be 1\,\um~wide, had an actual width of $\sim$970\,\nm. The inner curves of nominally zero-radius corners were observed to have an inner radius of $\sim$60\,nm. We ascribe this curvature to imperfect fabrication resolution.

The devices were tested in a dipstick probe in liquid helium with a base temperature of 4.2\,K. The temperature of the sample stage was controlled by a heater element. We tested the devices by ramping a low-noise home-built current supply with a programmed ramp from 0 to 500$\,\mu\mathrm{A}$ and back down to 0$\,\mu\mathrm{A}$ with a 100\,Hz repetition rate and measuring the resulting voltage over the wire. We determined the critical current by recording the current at which a threshold voltage across the device was passed during the positive portion of the current ramp. All results presented here were measured on a single chip and in a single testing session.

\begin{figure}
\includegraphics{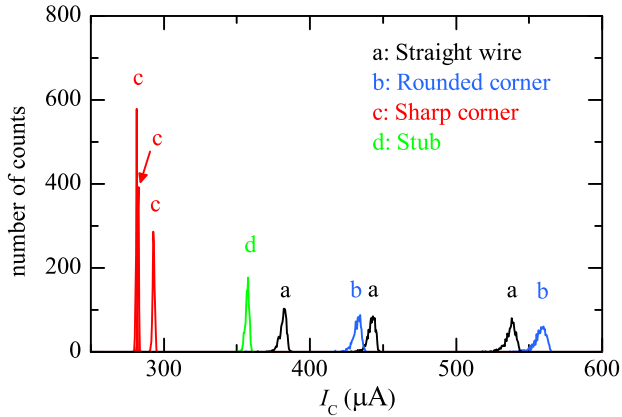}
\caption{Histograms of 1000 measured critical currents each, for 9 distinct devices: (a) straight lines (3 devices) 1-\um~wide; (b) corners with optimally designed inner corner curves (2 devices, straight segments were also 1-\um~wide); (c) corners as in (b) but in which the inner corner was fabricated to be as sharp as possible (3 devices); and (d) lines with a wider region (stub) extending 1 \um~to one side (1 device). The bin width is 0.5$\,\mu\mathrm{A}$.}
\label{histograms}
\end{figure}

Fig.~\ref{histograms} shows the resulting histograms from the critical current measurements at 4.2\,K. We recorded 1000 successive critical currents for each device. The straight wires (a) and wires with an optimally rounded inner corner (b) show averaged critical currents ranging from 382\,\uA~to 560\,\uA. All three devices with a sharp inner corner (c) exhibit a lower critical current between 280\,\uA~and 292\,\uA. Note that two of the sharp-corner histograms are overlapping at 280\,\uA. The device with a straight wire with a stub (d) has a critical current of 358\,\uA. The devices with sharp corners and the device with a stub all exhibit a narrower distribution of critical currents over their 1000 measurements, which can be seen from the height of the peak in the histogram.

We interpret the small variation in critical currents between the three different wires with a sharp corner (c) as an indication that indeed the corner is the dominant weakest link. For the straight (a) and rounded (b) wires there is no natural point for the switching to nucleate. Therefore, the critical current is sensitive to possible inhomogeneities. For example, on inspection we discovered that occasionally the wires contain a pinhole of $\sim$60-nm-diameter. These were absent in the devices with the highest critical current. The results for the wires with a sharp corner were independent of the presence of a pinhole. We emphasize that wire-to-wire variations are expected to be more likely to occur for uniform wires than for wires with a clearly defined weakest link.

We compared the measured values of the critical current to the expected critical pair-breaking current. Assuming a uniform current distribution, this current density is temperature dependent, and all material-dependent parameters are captured in a scaling parameter described by Romijn et al.\cite{Romijn}. Using the experimentally determined values for $T_\mathrm{c}=9.5$\,K and $R_\mathrm{s} = 217\,\Omega$, the only unknown in this equation is the diffusion constant $D$. The value of $D$ is unknown for our film, but if we use the value $D = 0.45\,\mathrm{cm^2/s}$ that was measured for a comparable NbN film\cite{Semenov}, we arrive at a critical current of $I_\mathrm{c} \approx 815\,\mu\mathrm{A}$ for our 1\,$\mu$m wide wire at $T$ = 4.2\,K. Given the uncertainty in the diffusion constant $D$, the observed critical current for our straight wires falls into the right range.

Using the results of Ref.~\cite{Clem}, for a corner with the observed inner radius of curvature of 60\,\nm, we expect $R\approx 0.33$ but observe a reduction of only 0.51 relative to the largest $I_{\rm c}$ measured. For the suddenly widening stub device, assuming zero inner radius of curvature of the corners, we expect $R\approx0.46$, but observe instead $R \approx 0.64$. This could in part be due to the finite minimal radius of curvature achievable in our fabrication process, but we note that the theory by Clem and Berggren does not take into account the superconducting coherence between the different current filaments. Taking this into account would require solving the Ginzburg-Landau equations or the more elaborate Usadel equations\cite{Usadel}. Nevertheless, the overall pattern of suppression is qualitatively clear: structures with sharp inner corners exhibit systematically lower values of $\Ic$ than straight wires and structures with rounded inner corners.

\begin{figure}
\includegraphics{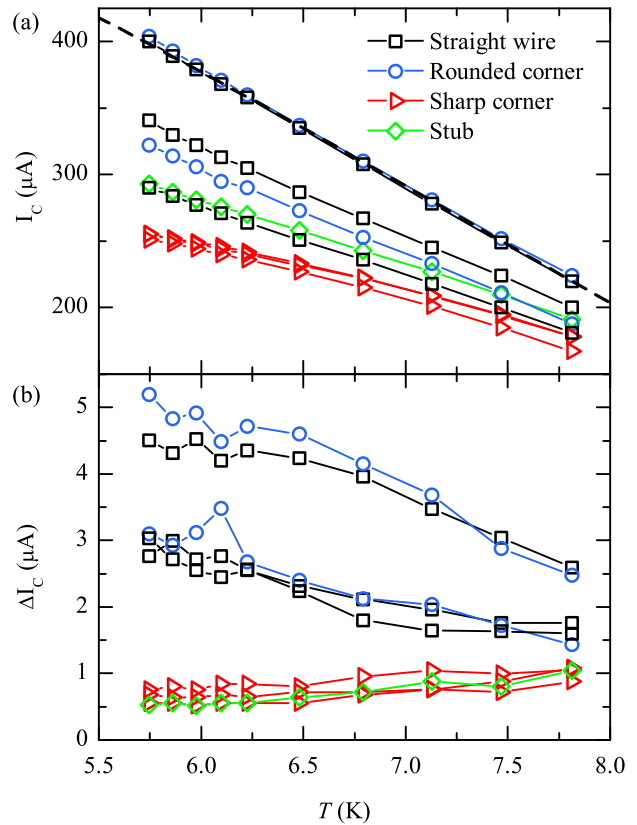}
\caption{Temperature dependence of (a) the critical current $\Ic$, and (b) the width of the critical current distribution $\Delta \Ic$ for straight wires (black squares), optimally rounded corners (blue circles), sharp corners (red triangles), and a wire with a stub (green diamonds). The dashed curve in (a) is a fit following Kupriyanov and Lukichev\cite{Kupriyanov}.}
\label{TempDependence}
\end{figure}

To provide further evidence for the impact of sharp corners on the critical current, in Fig.~\ref{TempDependence} we plot the temperature dependence of the critical current and of the width of the distribution of the critical current between 5.5\,K and 8\,K. For even higher temperatures, the current-voltage curves no longer exhibit a sharp transition from the superconducting to the normal state. At each temperature, histograms like the ones in Fig.~\ref{histograms} were taken with 1000 current ramps for each wire. The critical current $\Ic$ was taken to be the median of the critical current distribution. The width of the critical current distribution $\Delta \Ic$ was defined as the difference between the third and first quartile.

As shown in Fig.~4, the wires containing a sharp corner exhibit a significantly decreased critical current and a decreased width of their distribution of critical current, as seen before in Fig.~3. For all wires, the critical current decreases with increasing temperature, as expected. The difference between the unaffected and reduced critical currents decreases as the temperature is increased towards $T_\mathrm{c}$. The dashed curve in Fig.~\ref{TempDependence}(a) is a fit following Kupriyanov and Lukichev\cite{Kupriyanov}. This fit gives us a critical temperature $T_\mathrm{c}$ = 11.1\,K, in contrast to $T_\mathrm{c}$ determined from low-bias $R(T)$ measurements of 9.5\,K. We systematically observe such a difference, also in other NbTiN devices. Recently Benfatto et al. discussed the effects of a broadened Berezinskii-Kosterlitz-Thouless transition in disordered superconductors, which might explain this observation\cite{Benfatto}.

Another difference between wires is the width of their critical-current distribution, as shown in Fig.~4(b). For straight wires, the critical-current variance is approximately proportional to the critical current over the temperature range studied, $\Delta \Ic/\Ic(T)$ is constant. However, the wires with sharp geometric features exhibit a qualitatively different temperature dependence of $\Delta \Ic$. First, the critical-current distribution is consistently narrower. Furthermore, the absolute width of their critical-current distribution varies little with temperature, while the critical current decreases with increasing temperature. We interpret this difference also as being due to the sharp features, noting that they provide a natural weak spot at which the wire will initiate its transition to the normal state. In contrast, in the straight and optimally rounded devices, a fluctuation at any point in the wire could cause it to switch, and so the switching effect is not localized.

Finally, we return to the stub device (Fig.~2(d)). It is counterintuitive that a wider cross-section would lead to a lower critical current. As shown in Fig.~3 however, the stub device is at the lower end of the critical currents. Moreover, in Fig.~4(a) the temperature dependence is deviating from that of the straight wires, analogous to that of the sharp corners. Similarly Fig.~4(b) shows that the width of the critical current distribution is narrow, again analogous to the sharp corners. These observations taken together are taken as evidence that the stub indeed creates a dominant weakest link.

In summary, we have shown a clear reduction of the critical current in devices containing a sharp geometric feature. It is shown that the reduction of the critical current can be avoided by the inclusion of an optimally rounded corner. We further note that even in cases when $W > \Lambda$, which we have not treated theoretically or experimentally, as long as the condition $\xi < W$ is still satisfied, one would expect these phenomena to be important. The avoidance of critical current reduction may have an immediate impact on the performance of superconducting single-photon detectors and parametric amplifiers based on superconducting nanowires.

\textit{Noted added in proof:} During review of this paper, we became aware of recent related experimental work by Henrich et al.\cite{Henrich}

We thank Jules van Oven for help with electron microscopy and Kristen Sunter for help drafting figures. H.H. acknowledges support from the Foundation for Fundamental Research on Matter (FOM), K.B. also acknowledges support from the Netherlands Organization for Scientific Research NWO and the National Science Foundation (ECCS-0823778), E.D. acknowledges financial support from Microkelvin (No. 228464, Capacities Specific Programme) and the Teradec program of NWO-RFBR. This research was supported in part by the U.S. Dept. of Energy, Office of Basic Energy Science, Division of Materials Sciences and Engineering, and was performed in part at the Ames Laboratory, which is operated for the U.S. Dept. of Energy by Iowa State University under Contract No. DE-AC02-07CH11358.

\end{document}